\documentclass[
    ,final            
  ]
  {aipproc}

\usepackage{graphicx,amsmath,amssymb,epsfig,slashed}

\layoutstyle{6x9}

\newcommand{\mbf}[1]{\mbox{\boldmath $#1$}}
\newcommand{\bk}{\mbf{k}}

\newcommand{\ocs}[4]{\overset{#2}{{}_{#3}\slashed{\cal #1}_{#4}}}
\newcommand{\os}[4]{\overset{#2}{{}_{#3}\slashed{#1}_{#4}}}

\newcommand{\oc}[3]{\overset{#2}{{\cal #1}_{#3}}}

\newcommand{\obZ}[3]{\overset{#1}{\langle {}_{#2}Z_{#3} \rangle}}

\begin{document}

\title{Inclusive One Jet Production With Multiple Interactions
in the Regge Limit of pQCD}

\classification{11.55.Jy,12.38.Cy,13.85.Ni}

\keywords      {BFKL, Multiple Interactions, Regge Limit, Inclusive Production}

\author{G. P. Vacca}{
  address={INFN - Sezione di Bologna, Dip. di Fisica, Via Irnerio 46, 40126
  Bologna - Italy}
}

%

\begin{abstract}
DIS on a two nucleon system in the regge limit is considered. In this
framework a review is given of a pQCD approach for the computation of the
corrections to the inclusive one jet production cross section at finite number
of colors and discuss the general results. 
\end{abstract}

\maketitle


\section{Introduction}

 After experiments at HERA put in evidence the possibility of
 having kinematical regions characterized by high density hadronic matter, more
 investigations have been undertaken with nuclei at RHIC. Signals of high
 density effects will appear also in proton proton collision at LHC in the
 deep forward region. Typically tagging a not too hard jet in such a situation
 one expects multiple interaction between the emitted jet and one proton,
 which are typically classified as higher twist effects but which may be not
 negligible, power suppression being compensated by the resummation of
 some large energy logs.
 Such a situation can be somewhat more easily understood in the framework of
 photon-deuton scattering considering, due to the particular kinematics, small
 $x$ resummation techniques developed in a series of works after the pioneering BFKL~\cite{BFKL}
 approach.
 The study of multiple interaction both in total or jet inclusive cross
 section is also of great theoretical interest. In particular the computation
 requires to include some ingredients which are also needed to restore the unitarity,
 badly spoiled by too crude approximations (as e.g. the violation of the Froissart bound). 
 In this analysis one also meets the need to discuss the AGK~\cite{AGK} rules
 in the context of pQCD which are deeply interconnected to the inclusive one
 jet analysis~\cite{BRBSV,BSV2}.  

 I will first review an approach developed to study the total cross section case and
 after that describe how to procede to compute the cross section when one
 gluon jet is fixed at some rapidity. The results obtained should be compared with other approaches~\cite{Braun,KT,JMK,BGV,KL}. 

\section{Total cross section}
The total cross section for deep inelastic scattering on a nucleus consisting of two
weakly bound nucleons is constructed from the imaginary part of the
corresponding amplitude.
Following the discussion in ~\cite{AGK}, there are
three contributions illustrated in the left part in Fig.1. They are usually referred to 
as 'diffractive cut' (Fig.1a), 'single cut' (Fig.1b), and 'double cut' 
(Fig.1c).
\begin{figure}
\includegraphics[height=.1\textheight]{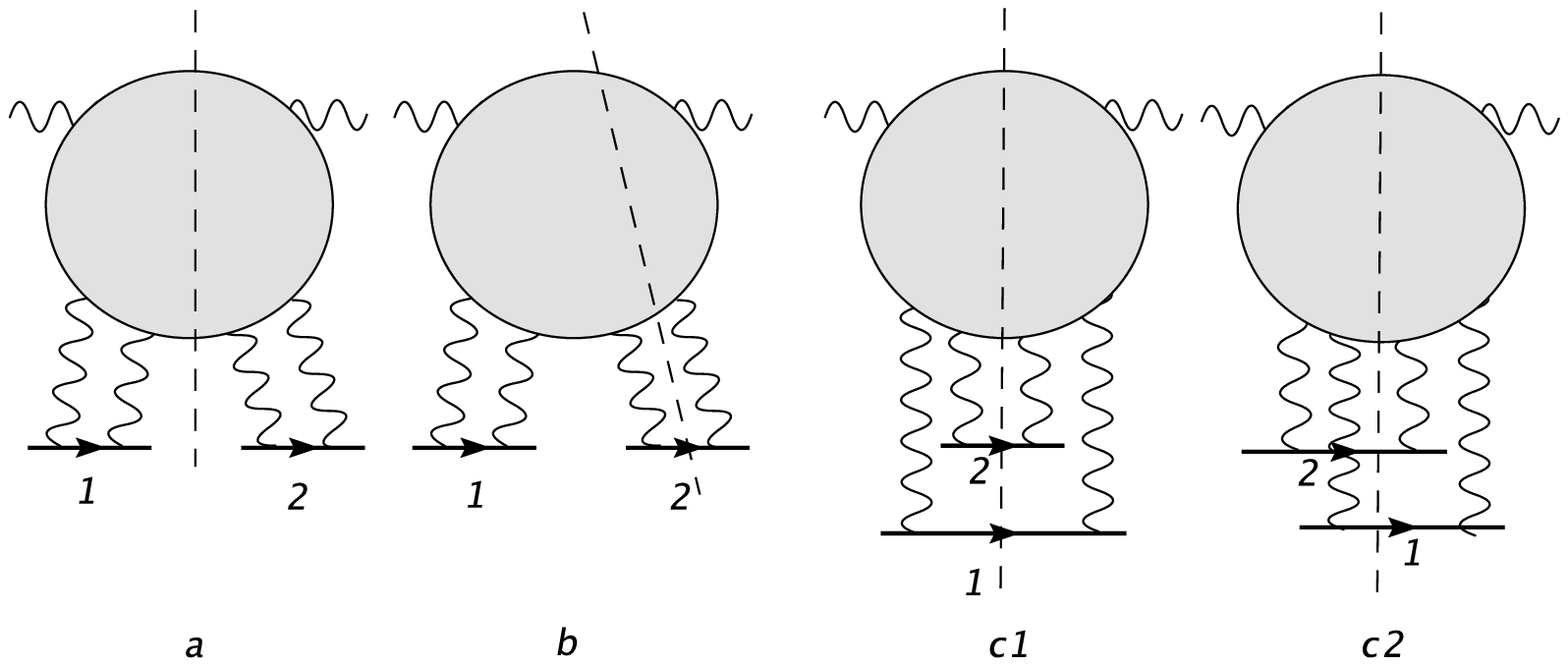}
\hspace{3cm}
\includegraphics[height=.1\textheight]{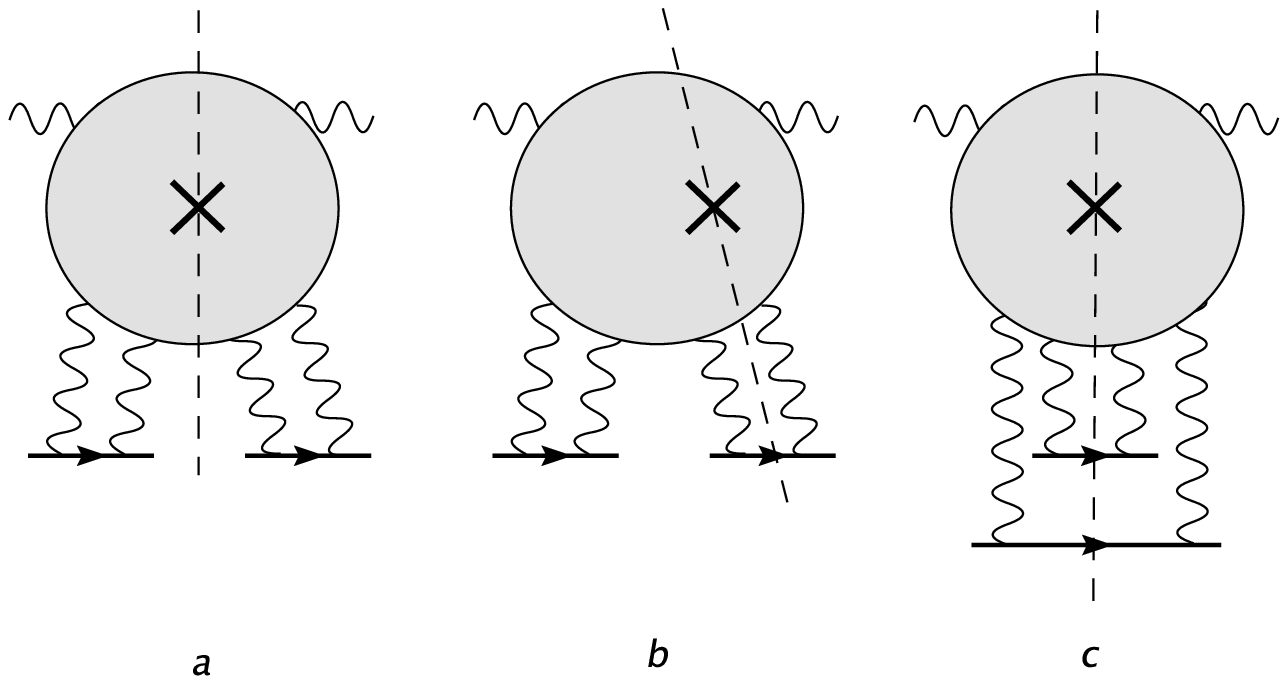}
\caption{Two cases: total cross section on the left and one jet inclusive on the
  right.
Different energy cuts: (a) diffractive cut, (b) single cut, (c) double cut.}
\end{figure} 
The total cross section is obtained from the sum of these terms plus some others obtained by
permutations among the nucleons.

In order to make explicit calculations it is convenient to introduce the light
cone momenta $q'$ and $p$ constructed from the external kinematical variables
and use the Sudakov decomposition for any momenta $k_i=\alpha_i q'+ \beta_i p
+ k_{i\,\perp}$ so that gluonic subamplitudes depend on the Sudakov
parameters $\beta_i, k_{i\,\perp}$. In our case subamplitudes depends on $4$
gluons and therefore on $\beta_1$,$\beta_2$ and $\beta_3$. The integration in
these variables always involve a path associated to the discontinuity related to
produced particles while the other two path are real. The gauge invariance leads to a good
behavior on the $\beta_i$ planes at infinity and deforming the contours one can
show that in all situations of Fig.1 after integration one obtains the same
function of transverse momenta $N_4 (\bk_{1\,\perp},\bk_{2\,\perp},\bk_{3\,\perp},\bk_{4\,\perp})$. 
This fact leads to the fullfillment of the AGK rules so that the in the high energy
limit the correction due to double interaction to the cross section are
negative and constructed from Fig.1 using different weigths: $-1=+1-4+2$.  
Another useful fact is that the single discontinuity amplitudes can be
constructed from the triple discontinuity ones~\cite{BSV2}. Therefore just
providing the correct phases one can define the total cross section from one
triple discontinuity calculation.

The triple discontinuity which is proportional to the total cross section is
defined by a set of coupled Bethe-Salpeter evolution equations~\cite{BW,BV} for
two, three and four reggeized gluon subamplitudes together with a set of
initial conditions. The solution of these equations can be decomposed in a
reducible reggeized part and a term satisfying Ward identities, characterized
by an effective $2$-to-$4$ conformal covariant transition vertex and a
$4$-gluon Green's function, which may be exactly constructed only in the large
$N_c$ approximation, when integrability appears. In the nuclei for large $N_c$
the Green functions factorize in two BFKL pomeron Green's functions and in the
M\"obius representation the effective vertex provides a triple pomeron
interaction~\cite{BV, BRV} which is mostly studied in the from given in the non linear BK equation~\cite{BK}.

\section{One jet inclusive cross section}
Let us know consider the one jet inclusive cross section.
When a jet is fixed in its rapidity and transverse momentum, among the
produced gluons in the generalized LLA approximation, one faces a new
situation which is depicted in the right part of Fig.1.  One can show that,
since the situation is less inclusive compared to the previous case, after the
integration in the $\beta_i$ Sudakov parameters the subamplitudes are no more
fully symmetric but still exhibit a symmetry on both sides of the cut. This means
that the three subamplitudes corresponding to Fig.1a-c on the right side are
different.
The inclusive cross section is therefore constructed summing the different
contribution each with a suitable phase factor and no factorization is
possible as before. The good news is that it is still possible to use the
triple discontinuities to reconstruct the three subamplitudes.

The task of computing the three different triple discontinuities is accomplished
generalizing the approach for the total cross section case.
Again one can write three coupled Bethe-Salpeter equations~\cite{BSV2} with suitable
initial conditions at the rapidity point where the jet is emitted. Such
conditions can be written in terms of the subamplitudes encountered in the total cross section analysis and of gluon production operators.

The subamplitudes $_iZ_n$ which solve the coupled system of equations,
$i$ denoting the position of the cut and $n$ the number of reggeized gluons,
are written as a sum of a \emph{reggeized part}
(constructed from a linear combination of solutions with a number of  reggeized gluons $< n$) and a \emph{irreducible part} which satisfies Ward identities, have correct symmetry properties
and define the effective vertices followed by Green's function evolution:
\begin{equation}
 {}_iZ_n  = {}_iZ_n^R + {}_iZ_n^I
\end{equation}

The analysis of the equations becomes non trivial already at the three gluon level.
Indeed complete reggeization breaks down and considering, due to azimuthal
symmetry in the jet momentum, an even function of the latter, one can show that new
terms beyond the reggeized ones do appear, nevertheless containing
effective $2$-to-$3$ transition verteces which nicely satisfy Ward
identities. These verteces , denoted $\os{\Gamma}{}{i}{3}$ in~\cite{BSV2},
are not present in total cross sections, in agreement with signature conservation. 

The four gluon subamplitudes can be analyzed following the decomposition above.
The reggeizing terms are of different kinds: there are terms similar to the
total cross section case, the jet being emitted at a runf of the BFKL pomeron ladder, wherein the presence of the jet is not disturbing reggeization. There is also a new term where the jet emission is taking place in the effective verteces $\os{\Gamma}{}{i}{3}$ followed by a rapidity evolution governed by three gluon pomeron states. Such Pomeron evolution is dual to the odderon evolution~\cite{BLV}.

Among the irreducible terms one finds that two of them are the ones naively expected, the jet being emitted again along a BFKL pomeron ladder or in a rung of the interacting $4$ reggeized gluons system corresponding to the Green's function evolution.
The remaining two terms are less trivial. One can be interpreted as characterized by jet emission from an effective production vertex, denoted $\ocs{V}{}{i}{4}$ in \cite{BSV2}. This vertex again has nice symmetry properties and satisfies Ward Identities,  which are crucial to pass eventually to a M\"obius representation~\cite{BLV2}.
Finally the last term is characterized by the jet emission from the effective vertex $\os{\Gamma}{}{i}{3}$, followed by an evolution of the reggeized gluons which at some point split by means of a new disconnected effective vertex in a $4$ gluon state. The latter $3\to4$ effective vertex satisfies clearly Ward identities and  has the same symmetry properties. It has been denoted by ${}_i\oc{W}{}{4}$ and its expression may be found in~\cite{BSV2}. We give here pictorially the structure of the solution:

\begin{equation}
  \label{eq:iZ4-total-pic}
  \begin{split}
    \obZ{}{i}{4} =
    \sum \bigg(
    \begin{minipage}{.9cm}
      \includegraphics[width=.9cm]{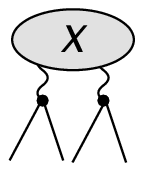}
    \end{minipage} +
    \begin{minipage}{.9cm}
      \includegraphics[width=.9cm]{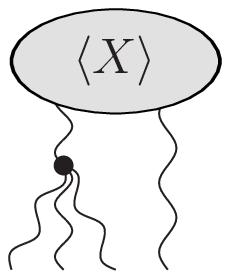}
    \end{minipage} +
    \begin{minipage}{.9cm}
      \includegraphics[width=.9cm]{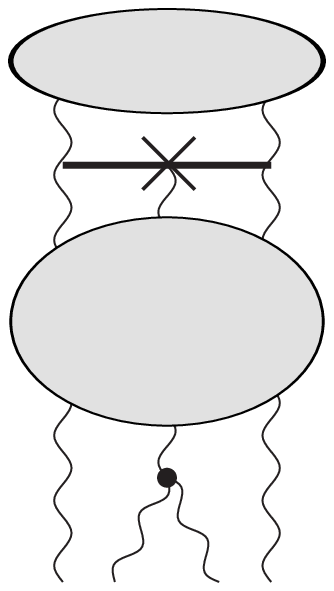}
    \end{minipage} +
    \begin{minipage}{1.3cm}
      \includegraphics[width=1.3cm]{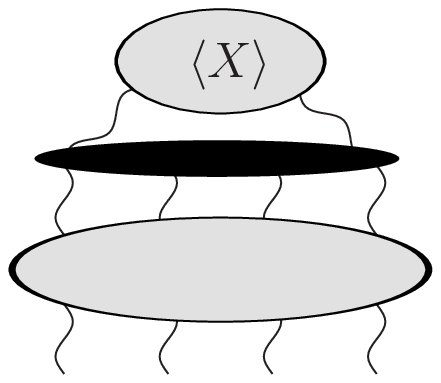}
    \end{minipage}
    \bigg) + 
    \begin{minipage}{1.3cm}
      \includegraphics[width=1.3cm]{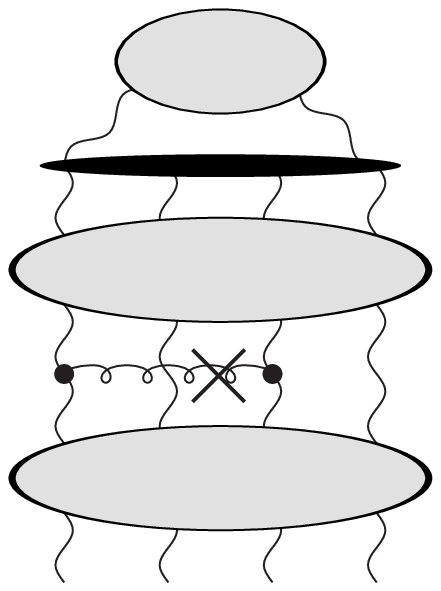}
    \end{minipage} +
    \begin{minipage}{1.3cm}
      \includegraphics[width=1.3cm]{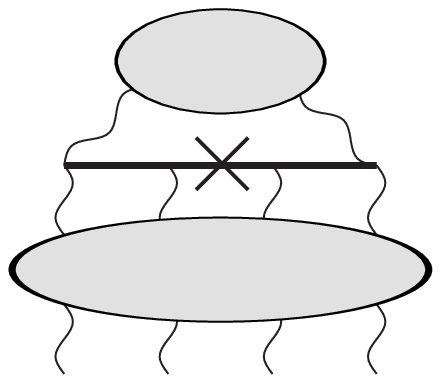}
    \end{minipage} +
    \begin{minipage}{1.3cm}
      \includegraphics[width=1.3cm]{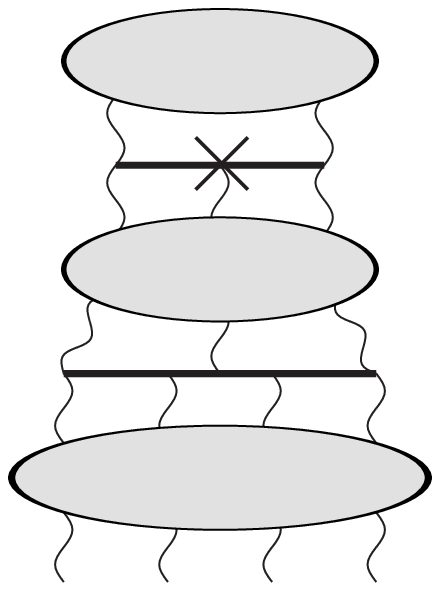}
    \end{minipage} \, .
  \end{split}
\end{equation}

We remind that the reggeized terms have a clear physical meaning, they play the role of higher order correlators inside the target.

At this stage the different single cut subamplitudes present a pattern of contribution which is similar to what has appeared in other computations, apart from a new term which was the latter described.
In order to obtain a physical one jet inclusive cross section the different subamplitudes should be  added, each multiplied with a suitable phase factor, as described in details in~\cite{BSV2}.
After that to compare explicitely the result with the other approaches and see if the latter term with the jet emitted at the $2\to3$ effective vertex is physically present in the one jet inclusive cross section, or is only relevant for the different multiplicities in the final state, one should pass  to the M\"obius representation. After that, as it was the case for the comparison to the dipole picture~\cite{Mueller,NZ}, one expectes huge simplifications in the long operatorial expressions.
This work will be carried on soon. 


\begin{theacknowledgments}
 The results reported were obtained in collaboration with J. Bartels
 and M. Salvadore. 
\end{theacknowledgments}



\end{document}